\documentclass[twocolumn,showpacs,showkeys,preprintnumbers,prb,superscriptaddress,floatfix]{revtex4-1}
\usepackage{amsmath,amssymb,mathrsfs}
\usepackage{graphicx}
\usepackage{bm}
\usepackage{subfigure}
\usepackage{hyperref}

\begin{document}
\title{$H$-$T$ Phase Diagram of Multi-component Superconductors with Frustrated Inter-component Couplings}
\author{Y. Takahashi}
\affiliation{International Center for Materials Nanoarchitectonics (WPI-MANA), National Institute for Materials Science, Tsukuba 305-0044, Japan}
\affiliation{Graduate School of Pure and Applied Sciences, University of Tsukuba, Tsukuba 305-8571, Japan}
\author{Z. Huang}
\affiliation{International Center for Materials Nanoarchitectonics (WPI-MANA), National Institute for Materials Science, Tsukuba 305-0044, Japan}
\affiliation{Graduate School of Pure and Applied Sciences, University of Tsukuba, Tsukuba 305-8571, Japan}
\author{X. Hu}
\affiliation{International Center for Materials Nanoarchitectonics (WPI-MANA), National Institute for Materials Science, Tsukuba 305-0044, Japan}
\affiliation{Graduate School of Pure and Applied Sciences, University of Tsukuba, Tsukuba 305-8571, Japan}
\date{\today}
\begin{abstract}
Multi-band superconductors in which frustrated inter-band couplings yield a time-reversal-symmetry breaking (TRSB)
state are investigated.
Stability condition for the TRSB state are derived based on the Bardeen-Cooper-Schrieffer (BCS) theory.
With the time-dependent Ginzburg-Landau (GL) method, vortex states are
investigated first at the vicinity of critical temperature $T_\text{c}$ where the GL theory is valid, and the results
are extended to compose the $H$-$T$ phase diagram.
When material parameters satisfy the condition that the nucleation field is slightly larger than the thermodynamic field ($H_\text{n} \gtrsim H_\text{tc}$) derived in a previous work (X. Hu and Z. Wang, Phys. Rev. B \textbf{85}, 064516 (2012)), an unconventional intermediate state characterized by clustering vortices appears.
Calculation of interface energy reveals that the clustering vortices are associated with  positive interface energy.
\end{abstract}
\maketitle
\section{Introduction}
Superconductivity associated with multiple condensations was discussed soon after the establishment of Bardeen-Cooper-Schrieffer (BCS) theory more than fifty years ago in view of compounds of transition metals.\cite{Suhl1959PRL,Kondo1963PTP,Leggett1966PTP} Later on, possible crucial differences between
multiband superconductors with frustrated Josephson-like couplings among condensates
and single-band superconductors attracted some attention.\cite{Agterberg1999PRB}
Recently, the discovery of iron-based high-temperature superconductors
\cite{Kamihara2008JACS, Ishida2009JPSJ} with at least three bands contributing to superconductivity, made it important to fully understand superconducting phenomena associated with the multibandness.

A straightforward extension of a single-band theory, as performed for two-band case is not sufficient for cases with more than three bands, since above three bands repulsive inter-band couplings may induce phase frustrations, and yields a bulk
state specified by phase differences among condensates different from 0 and $\pi$ and hence time-reversal symmetry breaking (TRSB, a term we shall use through the present work).\cite{Agterberg1999PRB,Stanev2010PRB,Yanagisawa2012JPSJ,Hu2012PRB}
A TRSB state was first proposed based on the BCS theory with $s$ pairing,\cite{Agterberg1999PRB,Stanev2010PRB} and later on other pairings such as  $s+id$ symmetry\cite{Lee2009PRL,Platt2012PRB} or $s+is$ symmetry\cite{Maiti2013PRB}
as well as junction structures\cite{Ng2009EPL,Lin2012PRB} have also been discussed.
Novel properties in collective excitations \cite{Ota2011PRB,Stanev2012PRB,Lin2012PRL,Maiti2013PRB} and chiral ground states were highlighted.\cite{Tanaka2010JPSJ,Dias2011SST,Yanagisawa2012JPSJ}

Within the scheme of multi-component Ginzburg-Landau (GL) theory on TRSB superconducting state, the thermodynamic field $H_\text{tc}$ and the nucleation field $H_\text{n}$ were analytically derived, and it was revealed that applying external magnetic field to such superconductor would generate interesting states, which might not be straightforwardly categorized to type-I or type-II by the GL parameter $\kappa = \lambda/\xi$.\cite{Hu2012PRB}
As a matter of fact, numerical calculations on magnetic properties exposed fractional vortices or unconventional
vortex states.\cite{Tanaka2010JPSJ,Garaud2011PRL,Carlstrom2011PRB,Yanagisawa2012JPSJ,Lin2012NJP,Babaev2012PhysC,Garaud2013PRB,Takahashi2013PhysC,Gillis2013arXiv}
However, a systematic study seems lacking at the moment of this writing.

In the present paper, we investigate magnetic properties of multicomponent superconductors with frustrated intercomponent couplings in the framework of GL theory, under the guidance from analytical results for the way to
choosing parameters in GL theory. The remaining part is organized as follows.
In Sec. II, we first derive the stability condition of the TRSB state based on multiband BCS theory.
In Sec. III, multicomponent GL formalism is introduced based on the multiband BCS theory.
In Sec. IV, vortex states in a TRSB superconductor are simulated with the time-dependent GL (TDGL) method,
and we find a vortex-cluster state. Based on this, we propose $H$-$T$ phase diagrams.
Finally, discussions and a summary follow in Sec. V and Sec. VI.

\section{Stability Condition of TRSB State}
We first discuss the stability condition of TRSB state based on the BCS theory.
The BCS Hamiltonian for a single band superconductor is extended to the multiband case,\cite{Suhl1959PRL}
\begin{gather}
H=\sum_{j\le N}\Big[
\sum_{\mathbf{k}_j,\sigma}\xi_{\mathbf{k}_j}c_{\mathbf{k}_j\sigma}^{\dagger}c_{\mathbf{k}_j\sigma}
-\sum_{\mathbf{k}_j,\mathbf{k}_j^\prime}V_{jj} c_{\mathbf{k}_j\uparrow}^{\dagger}c_{-\mathbf{k}_j\downarrow}^{\dagger}c_{-\mathbf{k}_j^\prime\downarrow}c_{\mathbf{k}_j^\prime\uparrow} \Big] \notag\\
-\sum_{j\neq l, j,l\le N}\sum_{\mathbf{k}_j,\mathbf{k}_l}V_{jl}c_{\mathbf{k}_j\uparrow}^{\dagger}c_{-\mathbf{k}_j\downarrow}^{\dagger}c_{-\mathbf{k}_l^\prime\downarrow}c_{\mathbf{k}_l^\prime\uparrow},
\end{gather}
with $N\geq3$.
The second and third terms correspond to the intra- and inter-band couplings, respectively.
The coupled self-consistent BCS gap equations for multiband superconductors are derived straightforwardly by a mean-field approach,\cite{Tinkham2004}
\begin{align}
\Delta_j=V_{jj}N_j\Delta_j\int_{-\hbar\omega_D}^{\hbar\omega_D}\frac{d\xi_j}{2E_j}\tanh\Big(\frac{E_j}{2k_\text{B} T}\Big) \notag\\
+\sum_{j\neq l}V_{jl}N_l\Delta_l\int_{-\hbar\omega_D}^{\hbar\omega_D}\frac{d\xi_l}{2E_l}\tanh\Big(\frac{E_l}{2k_\text{B} T}\Big),
\label{GapEq}
\end{align}
with $\Delta_j=-\sum_{\mathbf{k}_j} V_{jj}\langle c_{-\mathbf{k}_j\downarrow}c_{\mathbf{k}_j\uparrow}\rangle -\sum_{j\neq l} \sum_{\mathbf{k}_l} V_{jl}\langle c_{-\mathbf{k}_l\downarrow}c_{\mathbf{k}_l\uparrow}\rangle$ and $E_j=\sqrt{\xi_j^2+|\Delta_j|^2}$;
here an identical cut-off energy $\hbar\omega_D$ is taken in all bands for simplicity.
The coupled BCS equations for a three-band case are rewritten,
\begin{align}
\left[
\begin{array}{ccc}
  g_{11}-f_1(E_1, T) & g_{12} & g_{13} \\
  g_{12} & g_{22}-f_2(E_2, T) & g_{23} \\
  g_{13} & g_{23} & g_{33}-f_3(E_3, T)
\end{array}
\right] \notag\\
\times\left[
\begin{array}{c}
  \Delta_1 \\
  \Delta_2 \\
  \Delta_3
\end{array}
\right]
=\left[
\begin{array}{c}
  0 \\
  0 \\
  0
\end{array}
\right],
\label{gaps}
\end{align}
where,
\begin{equation}
f_j(E_j, T)=N_j\int_{-\hbar\omega_D}^{\hbar\omega_D}\frac{d\xi_j}{2E_j}\tanh\left(\frac{E_j}{2k_\text{B} T}\right),
\label{IntFunc}
\end{equation}
and $g_{jl}=[\mathbf{V}^{-1}]_{jl}$ with
\begin{equation}
\mathbf{V}=\left(
\begin{array}{ccc}
  V_{11} & V_{12} & V_{13}\\
  V_{12} & V_{22} & V_{23}\\
  V_{13} & V_{23} & V_{33}
\end{array}
\right). \notag
\end{equation}

Let us focus on the solution of Eq.~(\ref{gaps}) with complex gap functions which specify a TRSB state, where a necessary condition $g_{12}g_{23}g_{13}>0$ should be satisfied. It is easy to see, for example by putting $\Delta_1$ real as always
possible, one has the following equation,
\begin{align}
\left[
\begin{array}{cc}
  g_{22}-f_2(E_2,T)  & g_{23} \\
  g_{23} & g_{33}-f_3(E_3, T)
\end{array}
\right]
\left[
\begin{array}{c}
  {\rm Im}(\Delta_2)\\
  {\rm Im}(\Delta_3)
\end{array}
\right]
=\left[
\begin{array}{c}
  0 \\
  0
\end{array}
\right].
\end{align}
For nontrivial solution, one obtains the relation on the absolute values of the gap functions:
\begin{equation}
\left[g_{22}-f_2(E_2,T)\right]
\left[g_{33}-f_3(E_3,T)\right]
=g^2_{23}.
\label{TRSBcond2}
\end{equation}
In the same way, one has two other similar relations.
The three relations then yield
\begin{equation}
\left[ g_{jj}-f_j(E_j, T) \right]^2 = \frac{g_{jl}^2g_{jn}^2}{g_{ln}^2},
\label{SqrtCond}
\end{equation}
where $j\neq l \neq n$.
Noticing that, in the matrix Eq.~(\ref{gaps}), all diagonal terms $g_{jj}-f_j$ take the same sign as seen from Eq.~(\ref{TRSBcond2}), and there is only one independent vector, one can show that $g_{jj}-f_j$ and $g_{jl}g_{jn}/g_{ln}$ have the same sign.
The relations Eq.~(\ref{SqrtCond}) therefore are rewritten as
\begin{equation}
f_j(E_j,T)=g_{jj}-\frac{g_{jl}g_{jn}}{g_{ln}}.
\label{GapEqTRSB}
\end{equation}
Interestingly, they take the same form as a single-band case, except for that the intraband coupling is renormalized by the interband ones.

There is only one independent vector in the matrix in Eq.~(\ref{gaps}),
\begin{equation}
\frac{\Delta_1}{g_{23}}+\frac{\Delta_2}{g_{13}}+\frac{\Delta_3}{g_{12}}=0,
\label{SingleVector}
\end{equation}
which actually permits one to have complex solutions for Eq.~(\ref{gaps}).
From Eq.~(\ref{SingleVector}), it is clear that to search the complex solution one needs to form a triangle by using the three segments $|\Delta_j(T)/g_{ln}|$, and therefore
\begin{equation}
\left | \frac{\Delta_j(T)}{g_{ln}}\right | + \left | \frac{\Delta_l(T)}{g_{jn}}\right | >\left |  \frac{\Delta_n(T)}{g_{jl}}\right |.
\label{TRSBcond}
\end{equation}
One of these three inequalities may be broken as temperature sweeps, and the system transfers to a time-reversal symmetry reserved (TRSR) state where phase differences between the order parameters (OPs) take trivial values, i.e. $0$ or $\pi$.
The transition from TRSB to TRSR states takes place where one of the three inequalities is replaced by an equality
\begin{equation}
\left | \frac{\Delta_j(T)}{g_{ln}}\right | = \left | \frac{\Delta_l(T)}{g_{jn}}\right | + \left | \frac{\Delta_n(T)}{g_{jl}}\right |.
\end{equation}

The critical temperature of a TRSB superconductor can be derived from Eq.~(\ref{GapEqTRSB}) by putting $\Delta_j=0$,
\begin{equation}
N_j\ln\frac{2e^\gamma\hbar\omega_{D_j}}{\pi k_\text{B} T_\mathrm{c}}=g_{jj}-\frac{g_{jl}g_{jn}}{g_{ln}},
\end{equation}
with the Euler constant $\gamma=0.577215\cdots$.
In the following discussion, we focus on multiband superconductors with a TRSB state as an equilibrium bulk state.

\section{Multicomponent GL Theory}
\subsection{Derivation of GL equations}
We concentrate on temperature sufficiently close to the critical temperature where the GL theory is applicable.
By expanding the coupled BCS equations in Eq.~(\ref{gaps}) near $T_\text{c}$, we obtain the GL equations without gradient terms,
\begin{align}
\left( g_{jj}-N_j\ln\frac{2e^\gamma\hbar\omega_D}{\pi k_\text{B} T}\right)\Delta_j +\frac{7\zeta(3)N_j}{16(\pi k_\text{B} T_\mathrm{c})^2}|\Delta_j|^2\Delta_j\notag\\
+g_{jl}\Delta_l+g_{jn}\Delta_n=0,
\label{mGLeq}
\end{align}
with $\zeta(3)$ the Riemann zeta-function.
With the conventional expression for the GL equations and taking the OPs as $\psi_j=|\psi_j|e^{i\phi_j}=\Delta_j$, \cite{Zhitomirsky2004PRB,Gurevich2007PhysC,Yanagisawa2012JPSJ,Hu2012PRB,Orlova2013PRB}
\begin{align}
f =&\sum_{j=1,2,3} \bigg[ \alpha_j \lvert\psi_j\rvert^2 + \frac{\beta_j}{2} \lvert\psi_j\rvert^4 + \frac{1}{2m_j} \Big\rvert ( \frac{\hbar}{i} \nabla - \frac{2e}{c} \mathbf{A})\psi_j \Big\lvert^2 \bigg] \notag \\
&- \sum_{j<k} \gamma_{jk}(\psi_j \psi_k^\ast+\text{c.c.}) + \frac{1}{8\pi}(\nabla \times \mathbf{A})^2,
\label{3CGL}
\end{align}
one has
\begin{align}
\alpha_j(T) = -\left[N_j\ln\frac{2e^\gamma\hbar\omega_D}{\pi k_\text{B} T}-g_{jj}\right],\notag\\
\beta_j = \frac{7\zeta(3)N_j}{16(\pi k_\text{B} T_\mathrm{c})^2},\ \gamma_{jl}=-g_{jl}.
\end{align}
$m_j$'s are to be given in the same way as a single-band superconductor case, neglecting
the cross terms between different bands as in other works.

\subsection{Magnetic Properties in TRSB Superconductors}
The thermodynamic field $H_\text{tc}$ and nucleation field $H_\text{n}$ in a TRSB superconductor were analytically derived in Ref.~[\onlinecite{Hu2012PRB}]. However, an analytic treatment on
vortex states in the TRSB superconductor is very difficult, where spatially intertwined amplitudes and phases of OPs are crucial.

For simplicity, we consider a case that the second and third bands are same, but different from the first band.
In this case, the two typical fields are given by,
\begin{equation}
H_\text{tc}=2\sqrt{\frac{2\pi}{\beta_1}}\bigg( -\alpha_1-\frac{\gamma_{12}\gamma_{13}}{\gamma_{23}}\bigg)
\bigg( \frac{1}{2}+\frac{r_\alpha^2}{r_\beta}\bigg)^{1/2},
\label{H_tc}
\end{equation}
\begin{equation}
H_\text{n} = 2\kappa_1\sqrt{\frac{2\pi}{\beta_1}}\bigg( -\alpha_1-\frac{\gamma_{12}\gamma_{13}}{\gamma_{23}}\bigg)
\frac{r_\alpha r_m+2+|r_\alpha r_m- 1|}{2 +1/r_\alpha r_m},
\label{H_n}
\end{equation}
with $r_\alpha=\alpha_{2,3}/\alpha_1$, $r_\beta=\beta_{2,3}/\beta_1$, $r_m=m_{2,3}/m_1$ and $\kappa_1=(m_1c/2e\hbar)\sqrt{\beta_1/2\pi}$ is a material dependent parameter for the first component.

As the simplest but nontrivial case, we take $r_\alpha=1$, $r_\beta=1$ and $\gamma_{12}=\gamma_{23}=\gamma_{13}$ corresponding to an isotropic bulk state, while sweep the mass ratio $r_m$ between 0 and 1.\cite{Note}
In this case, the ratio between nucleation and thermodynamic fields is given by
\begin{equation}
\rho= H_\text{n}/H_\text{tc}
=\kappa_1\frac{3r_m}{2r_m+1}\sqrt{\frac{2}{3}}.
\label{Hratio2}
\end{equation}

As in a single component superconductor, magnetic responses of multicomponent superconductors change drastically across $\rho=H_\text{n}/H_\text{tc}=1$, which corresponds to a characteristic value of $\kappa_1$
\begin{equation}
\kappa_1^\ast=\frac{2r_m+1}{3r_m}\sqrt{\frac{3}{2}}.
\label{Kappa1star}
\end{equation}
Hereafter we perform our numerical study by varying the value of $\kappa_1$.
For convenience, dimensionless quantities are introduced.\cite{Lin2011PRB}
Units of length and energy are given by $\lambda_{10}=\sqrt{m_1c^2\beta_1/8\pi|\alpha_{10}|e^2}$ and $H_\text{1c}^2(0)=4\pi|\alpha_{10}|^2/\beta_1$, which are defined in the individual ($\gamma_{jk}=0$) first-component at $T=0$ with $\alpha_{10}=\alpha_1(T=0)$.

\subsection{Vortex States in TRSB superconductors}
\begin{figure*}
\includegraphics[width=\textwidth]{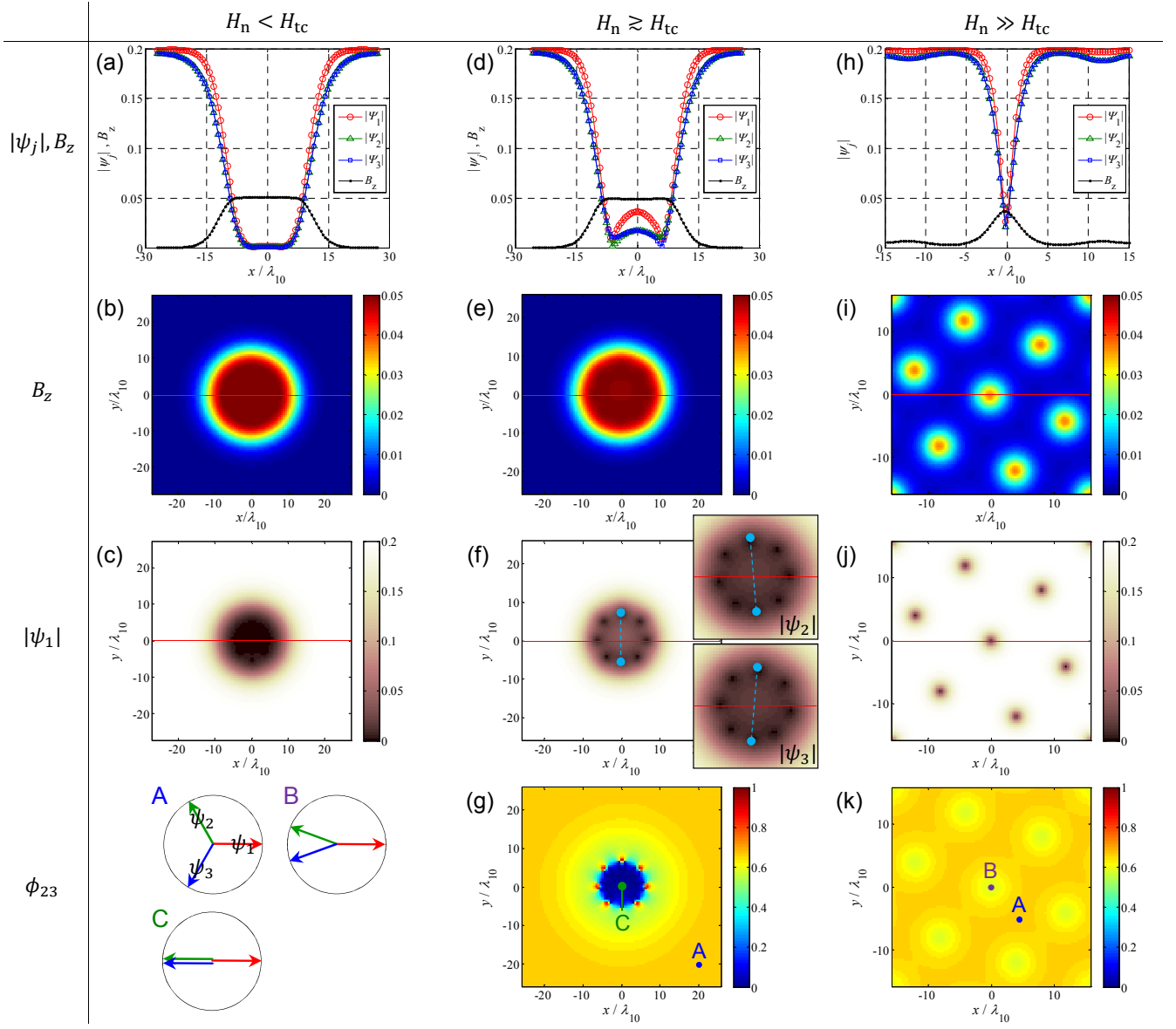}
\caption{(Color online) Typical vortex configurations solved by the TDGL method for (a-c) $H_\text{n}<H_\text{tc}$, (d-g) $H_\text{n}\gtrsim H_\text{tc}$ and (h-k) $H_\text{n}\gg H_\text{tc}$, for
$\kappa_1=2.0$, $\kappa_1=2.25$ and $\kappa_1=6.0$, respectively, with
the mass ratio $r_m=0.3$, interband couplings $\gamma_{12}=\gamma_{23}=\gamma_{13}=-0.3|\alpha_{10}|$ and temperature $T=0.97T_\text{c}$.
Panels (a, d, h) are spatial profiles of $\lvert\psi_1\rvert$, $\lvert\psi_2\rvert$, $\lvert\psi_3\rvert$ and $B_z$ along $y=0$ (red line) on the other panels.
Panels (b, e, i) denote magnetic induction $B_z$.
Panels (c, f, j) denote the amplitude of order parameter of the first-component $\lvert\psi_1\rvert$.
Panels (g, k) denote phase difference between the second and third components $\phi_{23}\equiv\phi_3-\phi_2$.
\label{Fig1}
}
\end{figure*}
Vortex states in a TRSB superconductor are studied based on the TDGL method adopting the \textit{magnetic periodic boundary condition} which confines fixed number of vortices $N$ in the simulation box.\cite{Doria1989PRB,Wang1991PRB,Kaper1995JCP}

The results shown below are for $N=8$, while we have confirmed the main conclusions remain valid using large systems. As an example we take $r_m=0.3$ which gives $\kappa_1^\ast\approx 2.12$.

Figure~\ref{Fig1} shows vortex configurations at three distinctive parameter regimes, where $H_\text{n}<H_\text{tc}$ ($\kappa_1<\kappa_1^\ast$), $H_\text{n}\gtrsim H_\text{tc}$ ($\kappa_1 \gtrsim \kappa_1^\ast$) and $H_\text{n}\gg H_\text{tc}$ ($\kappa_1 \gg \kappa_1^\ast$).
In Fig.~\ref{Fig1}(a-c) where $\kappa_1<\kappa_1^\ast$, we observe a typical phase separation between superconductivity and normal states which represents a type-I superconductor.
This result is the same as the conventional single-component superconductors.
As seen in Fig.~\ref{Fig1}(a), the OPs show different recovery lengths from normal to bulk superconductivity region due to anisotropic mass ratio $r_m$.

For $\kappa_1\gtrsim\kappa_1^\ast$, we find an unconventional vortex state.
Figures~\ref{Fig1}(d-g) show a vortex cluster, where multiple vortices ($N=8$) form a domain of vortex bundle.
As seen in Fig.~\ref{Fig1}(d), the OPs have finite values inside the domain region which is clearly different from Fig.~\ref{Fig1}(a).
As seen in Fig.~\ref{Fig1}(g), phase differences inside the cluster are either zero or $\pi$ indicating a TRSR state, while the bulk region keeps a TRSB state ($\phi_{jk}\equiv\phi_k-\phi_j=2\pi/3$).
This phase separation between TRSB and TRSR states is essential for the stability of vortex cluster.
It is noted that for the vortices locating at the phase boundary the vortex cores are not overlapping for the three components.

For $\kappa_1\gg\kappa_1^\ast$, Figs.~\ref{Fig1}(h-k) show typical triangular vortex lattice configurations which represent a type-II superconductor.
 In this parameter regime, OP phase differences at the vortex core are slightly modulated from bulk values, and no TRSR domain can be observed.

It is noted that when the system is totally isotropic, namely $r_m=1$, the vortex-cluster state does not appear, where modulation in amplitudes and phases in OPs are decoupled as discussed in Ref.[\onlinecite{Hu2012PRB}].

\subsection{Interface energy in TRSB superconductors}
\begin{figure}
\includegraphics[width=6cm,clip]{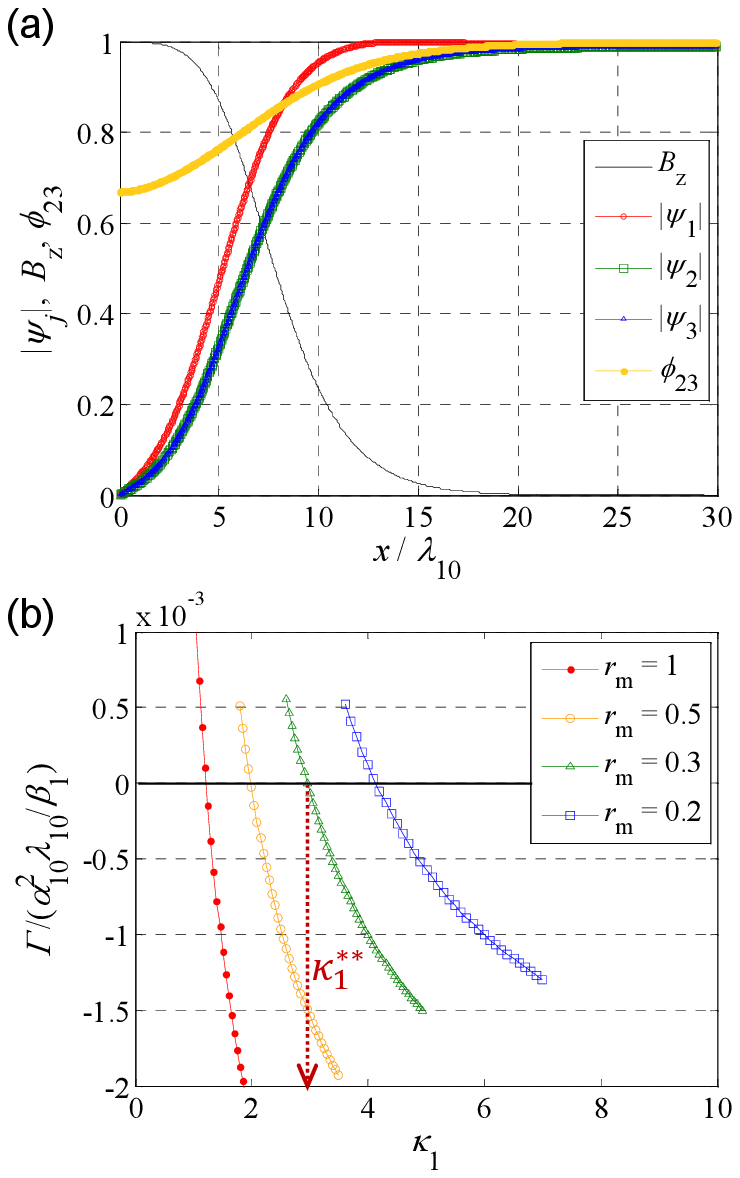}
\centering
\caption{(Color online) (a) TDGL calculation results of spatial variation for $\lvert\psi_j\rvert$, $\phi_{23}$ and $B_z$, which are normalized by bulk values $\lvert \psi_{j0}\rvert$, $\phi_{23}=2\pi/3$ and the thermodynamic field $H_\text{tc}$, respectively.
(b) $\kappa_1$ dependence of the interface energy $\Gamma$ in TRSB superconductors.
$\Gamma$ is normalized by the thermodynamic field for the first component without inter-component coupling.}
\label{Fig2}
\end{figure}

In a single component superconductor, the GL parameter $\kappa=1/\sqrt{2}$ given by the condition $H_\text{n}=H_\text{tc}$ coincides with that where sign change in interface energy takes place,\cite{Jacobs1979PRB,Lukyanchuk2001PRB} which dichotomizes a superconductor into type-I or type-II.
Therefore, it is interesting to evaluate interface energy in the TRSB superconductor where the vortex-cluster state appears, and thus the simple classification of type-I and type-II superconductor does not apply.
For this purpose, we calculate the excess Gibbs free energy in a one-dimensional system,\cite{Tinkham2004,Geyer2010PRB}
\begin{equation}
\Gamma = \int^{+\infty}_{0}(g_\text{sH}-g_\text{s0})dx,
\end{equation}
where, $g_\text{sH}$ and $g_\text{s0}$ are energy density with and without applied fields, respectively.

The boundary conditions are given,
\begin{align*}
\lvert \psi_j \rvert&=0 \ \text{and}\ B(x)=H_\text{tc} &\text{for } x\rightarrow 0 \notag \\
\lvert \psi_j \rvert&=\lvert\psi_{j0}\rvert \ \text{and}\ B(x)=0 &\text{for } x\rightarrow +\infty
\end{align*}
where $|\psi_{j0}|$'s are bulk values of OPs in each component, and $H_\text{tc}$ is the thermodynamic field of TRSB superconductor in Eq.~(\ref{H_tc}).

Typical interface structure is shown in Fig.~\ref{Fig2}(a) for $r_m=0.3$ and $\kappa_1=3$.
Figure~\ref{Fig2}(b) shows $\kappa_1$ dependence of the interface energy $\Gamma$ for several
typical values of mass ratio $r_m$.
Numerical errors are negligible in these plots.
The interface energy decreases monotonically with increase of $\kappa_1$ and changes its sign at $\kappa_1^{\ast\ast}$.
Therefore, in a TRSB superconductor, there are two threshold values of $\kappa_1$,
for example, $\kappa_1^{\ast\ast}\approx3.0$ and $\kappa_1^{\ast}\approx2.18$ for $r_m=0.3$, which
makes it much different from a TRSR superconductor.

\begin{figure}
\includegraphics[width=6cm,clip]{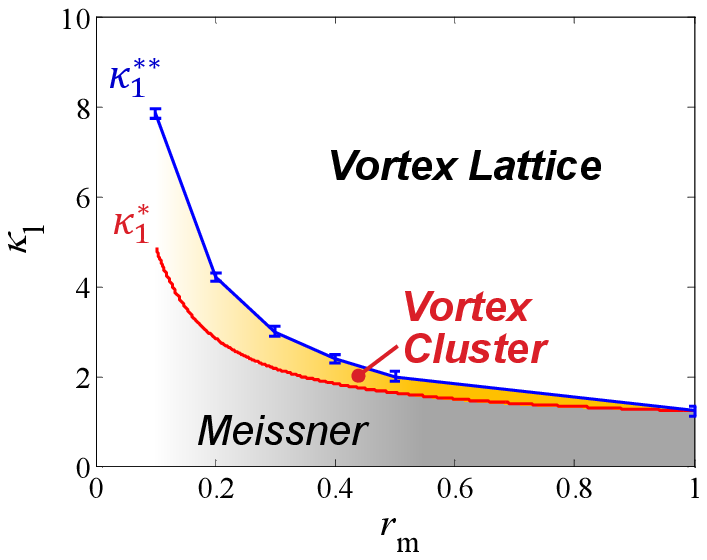}
\centering
\caption{(Color online) Phase diagram for vortex state in the TRSB superconductor in terms of $r_m$ and $\kappa_1$
including the Meissner, vortex cluster and vortex lattice phases.
See text for definitions.}
\label{Fig3}
\end{figure}
Based on the above results, a phase diagram is constructed in Fig.~\ref{Fig3}, with two phase boundaries
$\kappa_1^{\ast}$ and $\kappa_1^{\ast\ast}$ which separate the Meissner phase, vortex-cluster phase and vortex lattice phase.
The two phase boundaries overlap at $r_m=1$, which is consistent with that no vortex-cluster state can be found in the isotropic system.

\subsection{Vortex State at $H\lesssim H_{n}$}\label{TRS_Trans}
\begin{figure}
\includegraphics[width=6cm,clip]{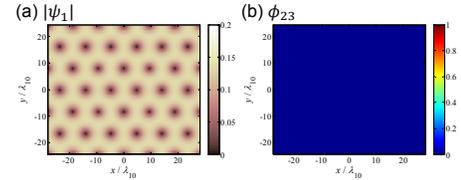}
\centering
\caption{(Color online) Vortex configurations with magnetic flux number $N=36$.
(a) Amplitude of $\psi_1$. (b) OP phase difference $\phi_{23}$.
Material parameters are same as in Fig.~\ref{Fig1}(d-g).}
\label{Fig4}
\end{figure}

Here we consider the field dependence of vortex states in the regime
$\kappa_1^\ast < \kappa_1 < \kappa_1^{\ast\ast}$ where a vortex cluster is observed. Variations of the system
upon sweeping applied magnetic fields is simulated by changing the number of vortices $N$ with fixed system size.
Figure~\ref{Fig4} shows the vortex configuration with the same material parameters and system size as those in Fig.~\ref{Fig1}(d-g) but with $N=36$.
A typical vortex lattice is observed, and phase difference are either 0 or $\pi$ as displayed
in Fig.~\ref{Fig4}(b) associated with a TRSR state in the whole system.
OPs are suppressed by the magnetic field in different ways in accordance with effective masses $m_j$,
which results in a breaking of the stability condition of TRSB state in Eq.~(\ref{TRSBcond}).
The magnetic-field-induced TRSB to TRSR transition is seen for $ \kappa_1>\kappa_1^\ast$.

\subsection{$H$-$T$ Phase Diagrams of TRSB superconductor}
\begin{figure*}
\includegraphics[width=\textwidth]{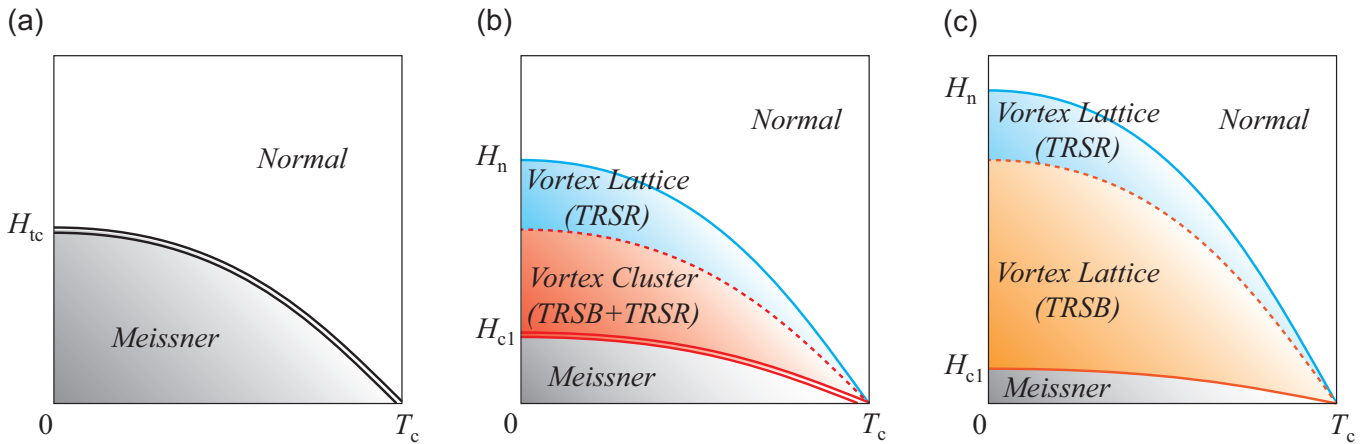}
\caption{(Color online) $H$-$T$ phase diagrams for multicomponent superconductors with frustrated intercomponent couplings.
Three diagrams are characterized by conditions: (a) $H_\text{n}<H_\text{tc}$ ($\kappa_1<\kappa_1^\ast$), (b)$H_\text{n}\gtrsim H_\text{tc}$ ($\kappa_1^\ast<\kappa_1<\kappa_1^{\ast\ast}$) and (c) $H_\text{n}\gg H_\text{tc}$ ($\kappa_1>\kappa_1^{\ast\ast}$).
The double and single lines represent first- and second-order transition, respectively.
The dashed line represents the TRSB-TRSR transition as discussed in Sec.~\ref{TRS_Trans}.
}
\label{Fig5}
\end{figure*}
In this section, we construct $H$-$T$ phase diagrams of multicomponent TRSB superconductor in the
 three regimes (a) $H_\text{n}<H_\text{tc} $ ($\kappa_1<\kappa_1^\ast$), (b) $H_\text{n}\gtrsim H_\text{tc} $ ($\kappa_1^\ast<\kappa_1<\kappa_1^{\ast\ast}$) and (c) $H_\text{n}\gg H_\text{tc} $ ($\kappa_1>\kappa_1^{\ast\ast}$).

In Fig.~\ref{Fig5}(a), the TRSB superconductor shows simply typical type-I property.
At high magnetic fields, superconductivity with a TRSB state is totally suppressed, and transfers to a normal state ($\lvert \psi_j \rvert=0$).
This is essentially the same $H$-$T$ phase diagram as a conventional single-component superconductor.
The phase transition between Meissner and normal states is unambiguously first order.

Figure~\ref{Fig5}(b) shows the novel $H$-$T$ phase diagram which includes the vortex-cluster state as an unconventional intermediate phase.
The vortex-cluster phase is located above the lower critical field $H_{c1}$ where vortices start to penetrate into a superconductor.
For the stronger magnetic field slightly below the nucleation field $H_\text{n}$, a conventional vortex lattice with TRSR state appears as shown in Fig.~\ref{Fig4}.

The phase transition between the vortex cluster and the Meissner states is likely not continuous.
The second-order transition at $H_\text{c1}$ in a type-II superconductor is understood by additional Gibbs free energy $\Delta G$ due to vortex penetration represented by\cite{Tinkham2004}
\begin{equation}
\Delta G=G_s\big\lvert_\text{first\ flux}-G_s\big\lvert_\text{no\ flux}=\frac{B}{\Phi_0}\epsilon_1 + \sum F_{ij} -\frac{BH}{4\pi},\notag
\end{equation}
where $\epsilon_1$ is vortex line energy and $F_{ij}=\frac{\Phi_0^2}{8\pi^2\lambda^2}K_0(\frac{r_{ij}}{\lambda})$ is vortex interaction energy with the zeroth Bessel function $K_0$.
For conventional case with repulsive vortex interaction, the energy cost is $\Delta G\approx 0$ because $B$ is much small inside the superconductor ($B\approx0$), and the interaction energy is also negligible ($F_{ij}\approx0$) as inter-vortex distance is large enough.
However when vortices form a cluster as observed in the TRSB superconductor, vortices penetrate into a superconductor feeling finite interaction energy $F_{ij}$, and consequently the system will see a finite energy jump $\Delta G$ which
corresponds naturally to a first-order transition.

Finally, Fig.~\ref{Fig5}(c) shows the $H$-$T$ phase diagram for $H_\text{n}\gg H_\text{tc} $ ($\kappa_1>\kappa_1^{\ast\ast}$).
Since the vortex lattice state is observed at magnetic fields $H_\text{c1}\leq H \leq H_\text{n}$, the phase diagram is essentially same as the single component case.
However, it is remarked that there are two regimes in terms of OP phase configurations.
For a low magnetic field regime, OP phases are locally modulated only in a vortex core, and the overall system preserves a TRSB state.
For a high magnetic field slightly below $H_\text{n}$, the system transfers to a TRSR state as seen in Fig.~\ref{Fig4}.
Between the two states, vortex configurations do not show obvious differences and inter-vortex distance changes proportionally to strength of applied magnetic field.

\section{Discussions}
Using a numerical approach based on multicomponent GL theory, we have revealed that, in a multicomponent superconductor with frustrated intercomponent couplings, a vortex-cluster state appears at an intermediate magnetic field regime between Meissner and vortex lattice states when the material parameters satisfy $H_\text{n}\gtrsim H_\text{tc}$.
While numerical results are shown explicitly for the case where the material parameter of the first component $\kappa_1$ and mass ratios between the components $r_m$ are varied whereas the other parameters relevant to a bulk value of OP are put the same,
the appearance of a vortex-cluster state is general for all possible parameters as far as the stability conditions of a TRSB state discussed in Sec. II are satisfied except the isotropic case.

The vortex-cluster state is expected to be observable by conventional vortex imaging methods.
It is also interesting to examine the behavior of magnetization around $H_\text{c1}$.
The magnetization curve will be different from either that of type-I or that of type-II superconductor.
Careful experiments are required and such unique magnetic behavior will also support the novelty of a TRSB superconductor.

It is appropriate to make some discussions on the nature of transition between the vortex cluster and vortex lattice (see discussions in Sec.~\ref{TRS_Trans} and Fig.~\ref{Fig5}(b)).
Since the spatial symmetry is different between the two states, a weak first-order transition is expected.
However, in the present work we could not find clear evidences for it since no thermodynamic quantity has been calculated directly.
On the other hand, the nature of the TRSB-TRSR transition in Fig. 5(c) is a more subtle issue.
This transition has only been discussed in absence of magnetic field (and thus without any vortex).\cite{Stanev2010PRB,Lin2012PRL}
While it was argued to be first order,\cite{Stanev2010PRB} a numerical analysis indicated a continuous transition.\cite{Lin2012PRL}
Therefore, the nature of TRSB-TRSR transition remains as an issue to be addressed in future works. 

Similar vortex-cluster states have been also reported in numerical studies based on the three-component GL theory.\cite{Carlstrom2011PRB,Garaud2011PRL}
It is mentioned in these studies that vortex cores of individual components do not overlap at the domain boundary, suggesting the existence of \textit{fractional vortices}.
In our study, similar vortex-cluster structure is observed as indicated on the panels in Fig.~\ref{Fig1}(f), where blue lines denote orientation of the vortex cluster.
However, separation of the cores is still unclear, and possibility of fractional vortices should be studied further.

While fractional vortices which appear at the boundary between two chiral TRSB superconductors were studied well,\cite{Yanagisawa2012JPSJ,Tanaka2010JPSJ,Garaud2011PRL}
those at a boundary between TRSB and TRSR states are also interesting, and deserve further study.

\section{Conclusion}
Magnetic properties of multiband superconductors with frustrated interband couplings are investigated.
The stability condition of the time-reversal-symmetry breaking state is derived based on Bardeen-Cooper-Schrieffer (BCS) theory for zero magnetic field.
Deriving multicomponent Ginzburg-Landau (GL) theory from the BCS theory, we have investigated response of the novel
 superconducting state to an external magnetic field.
When parameters satisfy the condition $H_\text{n} \gtrsim H_\text{tc}$, with $H_\text{n}$ the nucleation field and $H_\text{tc}$ the thermodynamic field, we have revealed the novel $H$-$T$ phase diagram including the unconventional vortex state, namely \textit{vortex cluster}, which cannot be categorized to either type-I or type-II.
The vortex cluster is associated with local domain separation between time-reversal symmetry broken and time-reversal symmetry reserved states, and it is expected to appear via a first order transition from the Meissner state.
We have studied the interface energy in a time-reversal-symmetry broken superconductor, and found that material parameters for sign change in the interface energy do not coincide with those for $H_\text{tc}=H_\text{n}$.
\ \\

\textit{Note added.} After completion of this work, we became aware of Ref.~\onlinecite{Wilson2013JPhys} which contains similar discussions for the stability condition of a time-reversal-symmetry breaking state based on BCS theory. 

\begin{acknowledgments}
This work was supported by the WPI initiative on Materials Nanoarchitectonics, and
partially by the Grant-in-Aid for Scientific Research (No. 25400385), MEXT of Japan.
\end{acknowledgments}

\appendix
\section{Magnetic properties of multiband TRSR superconductors}
We here discuss a multicomponent superconductor with intercomponent couplings \textit{unfrustrated}, i.e. $\gamma_{12}\gamma_{23}\gamma_{13}>0$, and analytically derive that it is similar to a single-component case
close to $T_\text{c}$.

The multicomponent GL equations are derived from the free energy density functional in Eq.~(\ref{3CGL}) by a variational method,
\begin{equation}
\label{MCGLeq}
\alpha_j\psi_j+\beta_j|\psi_j|^2\psi_j+\frac{1}{2m_j}\left( \frac{\hbar}{i}\nabla-\frac{2e}{c}\mathbf{A}\right)^2\psi_j
-\gamma_{jl}\psi_l - \gamma_{jn}\psi_n=0,
\end{equation}
and for supercurrents,
\begin{equation}
\frac{c}{4\pi}\nabla\times\nabla\times\mathbf{A}=\sum_j\frac{2e\hbar}{m_j}|\psi_j|^2\left(\nabla \phi_j-\frac{2\pi}{\Phi_0}\mathbf{A}\right),
\label{MCGL_Eq2R}
\end{equation}
with $j,l,n$ for indices of components and $\phi_j$ for phase of the OP.

Around the critical temperature, the order parameters are given by the linearized version of Eq.~(\ref{MCGLeq}).
\begin{equation}\label{LinGL}
    \begin{bmatrix}
        \alpha_1 & -\gamma_{12} & -\gamma_{13} \\
        -\gamma_{12} & \alpha_2 & -\gamma_{23} \\
        -\gamma_{13} & -\gamma_{23} & \alpha_3
    \end{bmatrix}
    \begin{bmatrix}
        \psi_1 \\
        \psi_2 \\
        \psi_3
    \end{bmatrix}
    =\mathbf{X}\cdot \mathbf{\Psi}=\mathbf{0}.
\end{equation}
The critical temperature $T_\text{c}$ is given when the determinant of $\mathbf{X}$ becomes zero.

To satisfy the condition that $\mathbf{X}$ is positive definite at $T> T_\text{c}$, all determinants of principal minors take non-negative values according to the Sylvester's criterion, namely $\alpha_j\geq0$ and $\alpha_j\alpha_l-\gamma_{jl}^2\geq0$.
For the case where $\mathbf{X}$ has a single zero eigenvalue at $T=T_\text{c}$ (in contrast to the
 case of two zero eigenvalues for the TRSB state\cite{Hu2012PRB}), one has
 $\sum_{j\neq l}\epsilon_j\epsilon_l=\sum_{j\neq l}(\alpha_j\alpha_l-\gamma_{jl}^2)>0$ with $\epsilon_j$ the eigenvalues of $\mathbf{X}$, which indicates that at least one term in the second summation should be
 positive (equivalently a single zero eigenvalue), for example $\alpha_2\alpha_3-\gamma_{23}^2>0$.

When $\mathbf{X}$ has two independent vectors at $T=T_\text{c}$, the ratios among order parameters for $T\leq T_\text{c}$ 
are given by the Cramer's rules from Eq.~(\ref{LinGL}),
\begin{gather}
\frac{\psi_2}{\psi_1}
=\frac{\alpha_1\alpha_3-\gamma_{13}^2}{\gamma_{12}\alpha_3+\gamma_{13}\gamma_{23}}
=\frac{\gamma_{12}\alpha_3+\gamma_{13}\gamma_{23}}{\alpha_2\alpha_3-\gamma_{23}^2},\notag\\
\frac{\psi_3}{\psi_1}
=\frac{\alpha_1\alpha_2-\gamma_{12}^2}{\gamma_{13}\alpha_2+\gamma_{12}\gamma_{23}}
=\frac{\gamma_{13}\alpha_2+\gamma_{12}\gamma_{23}}{\alpha_2\alpha_3-\gamma_{23}^2},
\end{gather}
where $\alpha_j\gamma_{ln}+\gamma_{jl}\gamma_{jn}\neq 0$ since
$\alpha_j\geq 0$ and $\gamma_{ln}\gamma_{jl}\gamma_{jn}>0$. 
It is noticed that the above relations indicate $\alpha_1\alpha_2-\gamma_{12}^2>0$
and  $\alpha_1\alpha_3-\gamma_{13}^2>0$. One then arrives at
\begin{gather}
\frac{\psi_2^2}{\psi_1^2}=\frac{\alpha_1\alpha_3-\gamma_{13}^2}{\alpha_2\alpha_3-\gamma_{23}^2},\notag\\
\frac{\psi_3^2}{\psi_1^2}=\frac{\alpha_1\alpha_2-\gamma_{12}^2}{\alpha_2\alpha_3-\gamma_{23}^2}.
\label{OPratio}
\end{gather}

For $T\leq T_\text{c}$, the OPs follow the coupled GL equations,
\begin{align}\label{GLeq}
   &\begin{bmatrix}
        \alpha_1+\beta_1\psi_1^2 & -\gamma_{12} & -\gamma_{13} \notag\\
        -\gamma_{12} & \alpha_2+\beta_2\psi_2^2 & -\gamma_{23} \notag\\
        -\gamma_{13} & -\gamma_{23} & \alpha_3+\beta_3\psi_3^2
    \end{bmatrix}
    \begin{bmatrix}
        \psi_1 \\
        \psi_2 \\
        \psi_3
    \end{bmatrix} \notag\\
    &=\mathbf{X}^\prime\cdot\mathbf{\Psi}=\mathbf{0}.
\end{align}
Since the determinant of $\mathbf{X}^\prime$ should be zero,one has the following OPs taking into account Eq.~(\ref{OPratio}),
\begin{gather}\label{OP}
\psi_1^2\approx\frac{-K_{23}\det\mathbf{X}}{\beta_1K_{23}^2+\beta_2K_{13}^2+\beta_3K_{12}^2},\notag\\
\psi_2^2\approx\frac{-K_{13}\det\mathbf{X}}{\beta_1K_{23}^2+\beta_2K_{13}^2+\beta_3K_{12}^2},\notag\\
\psi_3^2\approx\frac{-K_{12}\det\mathbf{X}}{\beta_1K_{23}^2+\beta_2K_{13}^2+\beta_3K_{12}^2},
\end{gather}
where $K_{jl}=\alpha_j\alpha_k-\gamma_{jl}^2$ up to $O(1-T/T_{\rm c})$.

Other quantities for a TRSR superconductor are straightforwardly available with the conventional approach.\cite{Tinkham2004}
The thermodynamic magnetic field is derived by free energy difference between superconductivity and normal state in absence of magnetic fields, namely $f_\text{n}-f_\text{sc}=\sum_{j=1,2,3}\alpha_j|\psi_j|^2+\beta_j|\psi_j|^4$,
\begin{equation}\label{Htc}
\frac{H_\text{tc}^2}{8\pi}=\frac{1}{2}\frac{(\det\textbf{X})^2}{\beta_1K_{23}^2+\beta_2K_{13}^2+\beta_3K_{12}^2}.
\end{equation}
\ 

In order to calculate the coherence length, we consider a one-dimensional system with the boundary condition that the order parameters recover from normal to bulk value, $|\psi_j|\rightarrow |\psi_{j0}|$ as $x\rightarrow +\infty$,
\begin{align}\label{common}
\frac{\hbar^2}{2m_1}\frac{\partial^2(\psi_1-\psi_{10})}{\partial x^2}=&\alpha_1(\psi_1-\psi_{10})+3\beta_1\psi_{10}^2(\psi_1-\psi_{10})\notag\\
&-\gamma_{12}(\psi_2-\psi_{20})-\gamma_{13}(\psi_3-\psi_{30}),\notag\\
\frac{\hbar^2}{2m_2}\frac{\partial^2(\psi_2-\psi_{20})}{\partial x^2}=&\alpha_2(\psi_2-\psi_{20})+3\beta_2\psi_{20}^2(\psi_2-\psi_{20})\notag\\
&-\gamma_{12}(\psi_1-\psi_{10})-\gamma_{23}(\psi_3-\psi_{30}),\notag\\
\frac{\hbar^2}{2m_3}\frac{\partial^2(\psi_3-\psi_{30})}{\partial x^2}=&\alpha_3(\psi_3-\psi_{30})+3\beta_3\psi_{30}^2(\psi_3-\psi_{30})\notag\\
&-\gamma_{23}(\psi_2-\psi_{20})-\gamma_{13}(\psi_1-\psi_{10}).
\end{align}
Taking $\psi_j-\psi_{j0}=a_j\exp(-\sqrt{2}x/\xi)$, the equations are rewritten,
\begin{widetext}
\begin{align}
\begin{bmatrix}
  \alpha_1 + 3\beta_1\psi_{10}^2 -\frac{\hbar^2\xi^{-2}}{m_1} & -\gamma_{12} & -\gamma_{13} \\
  -\gamma_{12} & \alpha_2 + 3\beta_2\psi_{20}^2 -\frac{\hbar^2\xi^{-2}}{m_2} & -\gamma_{23} \\
  -\gamma_{13} & -\gamma_{23} & \alpha_3 + 3\beta_3\psi_{30}^2 -\frac{\hbar^2\xi^{-2}}{m_3} \\
\end{bmatrix}\notag
\cdot
\begin{bmatrix}
  a_1 \\
  a_2 \\
  a_3 \\
\end{bmatrix}
=\mathbf{Y}\cdot\mathbf{a}=0.
\end{align}
\end{widetext}
The coherence length is then obtained from $\det\mathbf{Y}=0$,
\begin{equation}\label{xi}
\xi^{-2} \approx \frac{2}{\hbar^2}\frac{-\det{\mathbf{X}}}{K_{23}/m_1+K_{13}/m_2+K_{12}/m_3}.
\end{equation}

The London penetration depth $\lambda$ is straightforwardly obtained from the GL equation for supercurrents in Eq.~(\ref{MCGL_Eq2R}),
\ 

\begin{align}
\lambda^{-2}&=\frac{4\pi(2e)^2}{c^2} \left(\frac{\lvert\psi_{1}\rvert^2}{m_1}+\frac{\lvert\psi_{2}\rvert^2}{m_2}+\frac{\lvert\psi_{3}\rvert^2}{m_3} \right) \notag\\
\approx &\frac{4\pi(2e)^2}{c^2} \left(\frac{K_{23}}{m_1}+\frac{K_{13}}{m_2}+\frac{K_{12}}{m_3} \right) \notag\\
&\times \frac{-\det\textbf{X}}{\beta_1K_{23}^2+\beta_2K_{13}^2+\beta_3K_{12}^2}.
\label{lambda}
\end{align}

Finally, the nucleation field is derived from the linearized GL equations in the presence of fields $H$.
Taking the gauge $A_y=Hx, A_x=0, A_z=0$, the OPs can be expressed $\psi_j=e^{ik_yy}e^{ik_zz}f(x)$, which yield similar forms to the Schr\"{o}dinger equation,
\begin{align}
-\frac{\hbar}{2m_j}f_j^{\prime\prime}+\frac{\hbar}{2m_j}\left( \frac{2\pi H}{\Phi_0} \right)^2(x-x_0)^2f_j \notag\\
=-\left( \alpha_j+\frac{\hbar^2k_z^2}{2m_j} \right)f_j +\gamma_{jl}f_l + \gamma_{jm}f_m,
\end{align}
where $x_0=k_y\Phi_0/2\pi H$.
Based on the lowest Landau level solution with $f_j=b_j\exp\left[ -\frac{1}{2}\frac{2\pi H}{\Phi_0}(x-x_0)^2\right]$, we obtain
\begin{widetext}
\begin{align}
\begin{bmatrix}
  \alpha_1 + \frac{\hbar^2}{2m_1}\left( k_z^2 + \frac{2\pi H}{\Phi_0}\right) & -\gamma_{12} & -\gamma_{13} \\
  -\gamma_{12} & \alpha_2 + \frac{\hbar^2}{2m_2}\left( k_z^2 + \frac{2\pi H}{\Phi_0}\right) & -\gamma_{23} \\
  -\gamma_{13} & -\gamma_{23} & \alpha_3 + \frac{\hbar^2}{2m_3}\left( k_z^2 + \frac{2\pi H}{\Phi_0}\right) \\
\end{bmatrix}\notag
\cdot
\begin{bmatrix}
  b_1 \\
  b_2 \\
  b_3 \\
\end{bmatrix}
=\mathbf{Z}\cdot\mathbf{b}=0.
\end{align}
\end{widetext}
\ \\
The nucleation field is derived with $\det\mathbf{Z}=0$ and $k_z=0$,
\ \\
\begin{equation}\label{Hn}
H_\text{n}\approx\frac{\Phi_0}{2\pi}\frac{2}{\hbar^2}\frac{-\det\mathbf{X}}{K_{23}/m_1+K_{13}/m_2+K_{12}/m_3}.
\end{equation}

With the coherence length and the penetration depth, the characteristic fields can be rewritten
as $H_\text{tc}=\Phi_0/2\sqrt{2}\pi\xi\lambda$
and $H_\text{n}=\Phi_0/2\pi\xi^2$.
It it obvious that $H_\text{n}/H_\text{tc}=\sqrt{2}\lambda/\xi=\sqrt{2}\kappa$, with the GL parameter $\kappa$.
Therefore, magnetic properties in a TRSR superconductor are essentially the same as a conventional single-component case.


\begin{thebibliography}{99}
\bibitem{Note}
To see magnetic response of the intercomponent phases, it is convenient to focus on the
 regime $r_m<1$ where the second and third component become smaller than the first component, and
 the TRSB state is suppressed by magnetic field, which cannot be seen easily with $r_m>1$
 and equivalent second and third component.
\end{thebibliography}
\end{document}